\documentclass[aps,pre,twocolumn,10pt,superscriptaddress,nofootinbib,balancelastpage]{revtex4-1} 
\usepackage[latin1]{inputenc}
\usepackage{amsmath,amssymb}
\usepackage{mathrsfs} 
\usepackage{siunitx}
\usepackage{braket}
\usepackage{calc}
\usepackage{tabularx}
\usepackage{graphicx}
\usepackage{dsfont}
\usepackage{color}
\usepackage{ifthen}
\usepackage{fancyhdr}
\usepackage{appendix}
\usepackage{seqsplit}
\usepackage{upgreek}
\usepackage{xcolor}
\usepackage{booktabs}
\usepackage{array}
\usepackage{bm}
\usepackage[normalem]{ulem}
\graphicspath{{./}{Figures/}}

\newcommand{\mtx}[1]{\bm{\mathsf{#1}}}
\newcommand{\gvec}[1]{\vec{\mathfrak{#1}}}

\newcommand{\HH}{\mtx{\mathcal H}}
\newcommand{\DD}{\mtx{\mathcal D}}
\newcommand{\RR}{\mtx{\mathcal R}}
\newcommand{\KK}{\mtx{K}}
\newcommand{\CC}{\mtx{C}}
\newcommand{\OO}{\mtx{\Omega}}

\allowdisplaybreaks

\usepackage{etoolbox}
\usepackage{hyperref}
\apptocmd{\UrlBreaks}{\do\/\do\-}{}{}
\usepackage[capitalise]{cleveref}

\newcommand{\Null}{\mathbf{0}}%
\newcommand{\Eins}{\mathds{1}}%
\newcommand{\dif}{\mathrm{d}}%

\newcommand{\Nabla}{\vec{\nabla}}%
\newcommand{\Laplace}{\boldsymbol{\triangle}}%
\newcommand{\norm}[1]{\lVert#1\rVert}%
\newcommand{\diag}{\operatorname{diag}}%

\newcolumntype{Y}{>{\centering\arraybackslash}X}%
\newcolumntype{Z}{>{\raggedright\arraybackslash}X}%

\newlength{\myl}%
\newcommand{\SUM}[2]{{\setlength{\myl}{\widthof{$\displaystyle\sum_{#1}^{#2}$}*\real{0.5}-\widthof{$\displaystyle\sum$}*\real{0.5}}\sum_{#1}^{#2}\;\hspace{-\the\myl}}}
\newcommand{\INT}[3]{\settowidth{\myl}{$\displaystyle\int_{#1}^{#2}$}{\int_{#1}^{#2}\;\;\;\hspace{-\the\myl}\dif #3}\,}
\newcommand{\TINT}[3]{\settowidth{\myl}{$\int_{#1}^{#2}$}{\int_{#1}^{#2}\!\ifthenelse{\equal{#1#2}{}}{}{\;\;\;\;\hspace{-\the\myl}}\dif #3}\,}%
\newcommand{\EINT}[3]{\settowidth{\myl}{$\int_{#1}^{#2}$}{\int_{#1}^{#2}\;\;\;\,\hspace{-\the\myl}\dif #3}\,}

\begin{document}
	
\title{Light-propelled microparticles based on symmetry-broken refractive index profiles}

\author{Julian Jeggle}
\thanks{These authors contributed equally to this work.}
\affiliation{Institute of Theoretical Physics, Center for Soft Nanoscience, University of M{\"u}nster, 48149 M{\"u}nster, Germany}

\author{Matthias R\"uschenbaum}
\thanks{These authors contributed equally to this work.}
\affiliation{Institute of Applied Physics, University of M{\"u}nster, 48149 M{\"u}nster, Germany}

\author{Adrian Paskert}
\affiliation{Department of Physics, RWTH Aachen University, 52074 Aachen, Germany}
\affiliation{DWI -- Leibniz Institute for Interactive Materials, 52074 Aachen, Germany}
\affiliation{Institute of Theoretical Physics, Center for Soft Nanoscience, University of M{\"u}nster, 48149 M{\"u}nster, Germany}

\author{Ivan Kalthoff}
\affiliation{Department of Physics, RWTH Aachen University, 52074 Aachen, Germany}
\affiliation{DWI -- Leibniz Institute for Interactive Materials, 52074 Aachen, Germany}
\affiliation{Institute of Theoretical Physics, Center for Soft Nanoscience, University of M{\"u}nster, 48149 M{\"u}nster, Germany}

\author{Elena Vinnemeier}
\affiliation{Institute of Applied Physics, University of M{\"u}nster, 48149 M{\"u}nster, Germany}
\affiliation{Institute of Physics, University of M{\"u}nster, 48149 M{\"u}nster, Germany}

\author{Jesco Sch\"onfelder}
\affiliation{Institute of Physical Chemistry, Center for Soft Nanoscience, University of M{\"u}nster, 48149 M{\"u}nster, Germany}

\author{J\"org Imbrock}
\affiliation{Institute of Applied Physics, University of M{\"u}nster, 48149 M{\"u}nster, Germany}

\author{Cornelia Denz}
\affiliation{Physikalisch-Technische Bundesanstalt, 38116 Braunschweig, Germany}

\author{Marcel Rey}
\email[Corresponding author: ]{mrey@uni-muenster.de}
\affiliation{Institute of Physical Chemistry, Center for Soft Nanoscience, University of M{\"u}nster, 48149 M{\"u}nster, Germany}
\email[Corresponding author: ]{mrey@uni-muenster.de}

\author{Raphael Wittkowski}
\email[Corresponding author: ]{rgwitt25@dwi.rwth-aachen.de}
\affiliation{Department of Physics, RWTH Aachen University, 52074 Aachen, Germany}
\affiliation{DWI -- Leibniz Institute for Interactive Materials, 52074 Aachen, Germany}
\affiliation{Institute of Theoretical Physics, Center for Soft Nanoscience, University of M{\"u}nster, 48149 M{\"u}nster, Germany}

\begin{abstract}		
Active colloidal microparticles require reliable actuation to sustain directed motion. Light-based propulsion is particularly attractive as it provides persistent energy supply and enables direct spatiotemporal control. Here, we introduce 3D-printable particles with symmetry-broken refractive index profiles (SBRIP particles) that achieve propulsion through direct momentum transfer from asymmetric light refraction. Internal refractive-index gradients provide optical symmetry breaking independent of external shape, fundamentally decoupling propulsion from particle geometry. Geometrically symmetry-broken particles with a homogeneous refractive index are another special case, where propulsion originates from refractive contrast at the boundary instead of within the particle. Unlike conventional systems relying on absorption or reflection, this transparency-based mechanism minimizes heating and mitigates shadowing in bulk suspensions. We present a theoretical framework for refractive propulsion as well as numerical simulations of the SBRIP particles using raytracing and the finite volume method. This is complemented by experiments, validating the momentum transfer mechanism using particles with geometric symmetry breaking. The high transparency of our particles ensures deep light penetration, enabling the realization of volumetric active matter. This opens pathways toward adaptive nonlinear optical materials where light-driven particle reorganization modulates the local refractive index, establishing a dynamic feedback loop between the optical field and the material structure.
\end{abstract}
\maketitle

\section{Introduction}
Systems of active colloidal microparticles, microscopic agents capable of self-propulsion in a supporting medium, have been subject to intense study in the last two decades as an archetypal instance of active matter \cite{Bechinger2016, teVrugt2025,Gompper2025}. Unlike their passive counterparts, these active systems are intrinsically out of thermodynamic equilibrium, serving as artificial analogues to natural microswimmers like bacteria and protozoa \cite{Bechinger2016,Marchetti2013,Romanczuk2012}. Beyond their fundamental physical interest, active particles are promising building blocks for future technologies. Their ability to self-assemble into ordered phases offers a route to ``intelligent'' materials with tunable properties, where the dynamic alteration of particle interactions allows these systems to form reconfigurable networks or adaptive structures that exhibit a complexity rivaling biological neural systems \cite{Mallory2018,Manoharan2015,Yu2018,Walther2020,Kaspar2021,Wang2024}. Simultaneously, individual autonomous swimmers are envisioned as microrobots for tasks ranging from targeted drug delivery to environmental remediation \cite{Nelson2010,Soler2014,Safdar2017,Xiao2025}. Related approaches harness biological activity by placing passive microstructures into bacterial baths, where bacterial motion drives gears or sailboat-like objects \cite{DiLeonardo2010,Pellicciotta2025}. To realize this potential experimentally, however, robust actuation strategies are required that can overcome the dominance of viscous forces at the microscale and supply the continuous energy needed for directed motion \cite{Bechinger2016}.

To meet this challenge, a variety of biomimetic propulsion mechanisms has been developed \cite{Ju2025}. A prominent class of experimental realizations consists of chemically powered micro- and nanomotors that harvest free energy through asymmetric catalytic reactions \cite{Sanchez2015, Ebbens2010}. These systems typically employ Janus particles to decompose a dissolved fuel, creating local concentration or electrical gradients that drive motion via self-diffusiophoresis and self-electrophoresis \cite{Brown2014,Moran2017,Vutukuri2017}, or alternatively through the mechanical recoil of evolved gas bubbles \cite{Sanchez2015,Campos2023}. However, these chemically propelled systems face intrinsic limitations: they require a specific chemical environment and depend on the continuous consumption of fuel, which inevitably depletes over time \cite{Ebbens2010}. Furthermore, the generated chemical gradients often induce complex hydrodynamic flows in the surrounding medium \cite{Bechinger2016,Illien2017}.

To overcome these fuel-related issues, alternative approaches driven by external fields have been established, such as acoustically driven swimmers \cite{Xu2017}, magnetic-field-actuated particles including helical swimmers, disks, rods, as well as biohybrid microrobots \cite{Ghosh2009,Palacci2013a,Yigit2019,Gardi2022,Magdanz2020,Gentili2025,Barroso2015} and electric-field-driven Janus colloids that self-propel via induced electrohydrodynamic flows under alternating current fields \cite{Baalen2025,Kesteren2023,Ketzetzi2025,Tanuku2023}. However, the local modulation of externally applied magnetic and electric fields remains technically challenging. As a result, spatial selectivity is limited and the independent addressing of individual particles remains difficult. This restriction complicates the initiation of adaptive interactions that are central to concepts such as intelligent active materials or material neural networks.

Among external driving mechanisms, optical actuation is particularly advantageous. It allows for a contact-free and continuous energy supply, while the high degree of spatio-temporal control over light fields enables precise manipulation of the particle dynamics \cite{Safdar2017,Rey2023}. 
Current illumination-based approaches are diverse, ranging from direct light-induced body deformations and phase transitions to light-mediated chemical reactions or phoretic effects \cite{Sipova2020,Schmidt2024,Safdar2017,Rey2023,Yang2023,Vardhan2025}.
They include photothermal propulsion based on light-induced heating and thermo- or photophoretic effects \cite{Khadka2018,Qian2013,Kuemmel2013,ButtinoniBKLBS2013,Truong2026,Heuthe2025,Knippenberg2025,Palagi2016}. Other widely studied systems exploit photocatalytic reactions, where illumination triggers chemical activity that drives propulsion and self-organization \cite{Palacci2013,Vutukuri2020,Wang2019,Boniface2024}. In these systems, light typically serves as an energy trigger to induce a secondary physical effect, such as a thermal or chemical gradient, that drives the motion.
Consequently, a major disadvantage of most such approaches is their reliance on light absorption. In bulk suspensions, this absorption leads to unwanted heating and causes "shadowing effects", leaving particles in the interior of the system insufficiently activated.

Alternatively, motion can be generated via direct optical momentum transfer by redirecting light \cite{Palima2013}.
Early demonstrations include refractive lightfoils that experience a transverse optical lift force under uniform illumination \cite{Swartzlander2011}, as well as light sailboats, which use specular reflection from angled micromirrors to generate lateral forces \cite{Buzas2012}. In both cases, propulsion is achieved by redirecting light at the particle boundary, such that the net force is set by the external shape and surface orientation, coupling propulsion to the particle's hydrodynamic properties and constraining independent optimization. Moreover, reflection-based approaches inherently redirect optical power away from the incident beam. More recent implementations employ plasmonic nanomotors \cite{Tanaka2020} and structured photonic platforms such as metavehicles \cite{Andren2021}, light-driven microdrones \cite{Wu2022}, metarotors \cite{Shanei2025}, untethered metaspinners \cite{Engay2025}, and geared metamachines \cite{Wang2025}, which rely on strong directional scattering from nanofabricated nanoantennas or metasurfaces to achieve actuation \cite{Shanei2022,Mitra2023}. 
Combined with the tunability of light fields, refractive propulsion paves the way for novel forms of adaptive nonlinear optical matter, where light-driven particle rearrangement creates an intensity-dependent optical response. By exploiting the feedback between particle reorganization and light propagation, these systems enable complex emergent behaviors, ranging from robust light self-trapping \cite{Man2013} to swarm intelligence \cite{Wensink2012,Yang2010,Cohen2014}, opening new avenues for photonic information processing.
These approaches, however, require complex cleanroom nanofabrication processes and precise integration of asymmetric nanoscale features. This technically demanding process constrains the accessible design space limiting particle shape to two-dimensional objects. Furthermore, these structures can still exhibit non-negligible thermal dissipation\cite{Wang2025,Shanei2025}, reintroducing the heating issues that momentum-transfer mechanisms aim to avoid.

\begin{figure}
    \centering
    \includegraphics[width=\linewidth]{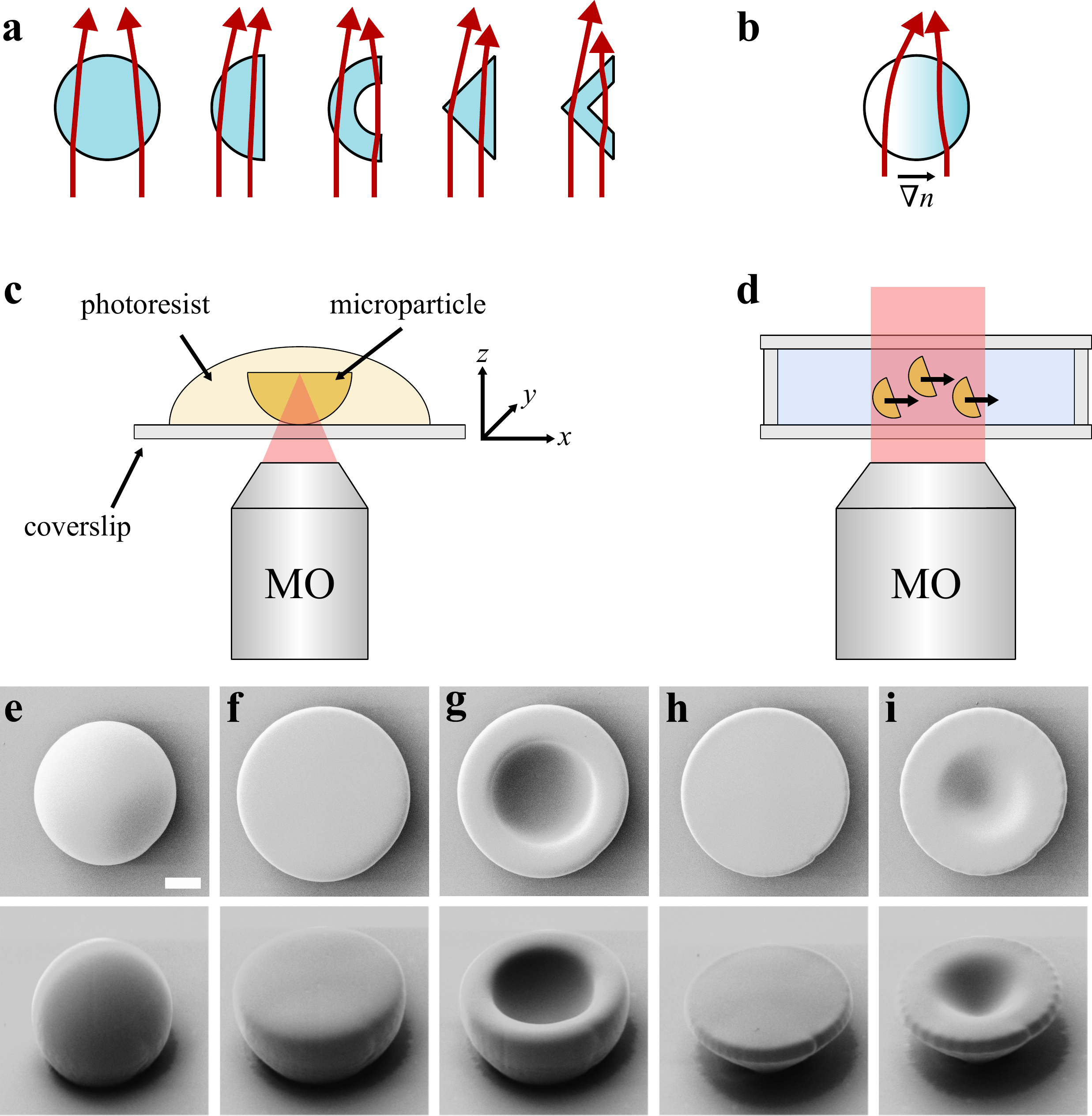}
    \caption{Design and fabrication of symmetry-broken refractive index profile (SBRIP) particles. \textbf{a} Cross sections of different particle shapes and corresponding light refraction: sphere, hemisphere, cap, cone, and cornet (from left to right). \textbf{b} Cross section of gradient-index (GRIN) sphere with refractive-index gradient $\nabla n$ ($n$ is represented by the saturation of the blue color) and light refraction. \textbf{c} Fabrication setup based on two-photon polymerization: a near-infrared femtosecond pulsed laser is tightly focused with a microscope objective (MO) into a droplet of photoresist (OrmoComp). The sample stage is then moved around the focus in three dimensions. \textbf{d} Propulsion setup: particles inside a sealed chamber filled with water are illuminated by an infrared laser from below, resulting in  propulsion. \textbf{e}-\textbf{i} Scanning electron microscopy (SEM) images of \textbf{e} spheres, \textbf{f} hemispheres, \textbf{g} caps, \textbf{h} cones, and \textbf{i} cornets. Top row: top view; bottom row: oblique view. Scale bar: \SI{2}{\micro \metre}.}
    \label{fig:01}
\end{figure}

Here, we introduce 3D-printable symmetry-broken refractive index profile (SBRIP) particles as a class of light-driven microparticles in which direct momentum transfer from asymmetric light refraction drives propulsion with minimal thermal losses \cite{Wittkowski2019,DenzSPIE2021,RueschenbaumSPIE2021,WittkowskiSPIE2021,RueschenbaumSPIE2023}. These particles feature a refractive index structure that leads to symmetry-broken light refraction, thus imparting a net momentum (see Fig.~\ref{fig:01}\textbf{a} and \textbf{b}). Since this propulsion mechanism does not rely on auxiliary flows or gradients being induced in the medium, this class of particles can be considered as an exact implementation of Active Brownian particles. Particles with a homogeneous refractive index but a symmetry-broken shape (see Fig.~\ref{fig:01}\textbf{a}) as well as particles with shape symmetry fabricated from a material with a refractive-index gradient (see Fig.~\ref{fig:01}\textbf{b}) emerge as special cases of this framework. The latter case is more challenging to implement, but allows a decoupling of particle activity from particle geometry and thus independent control over hydrodynamic properties and propulsion.  Importantly, 3D printing provides a versatile fabrication route for realizing such volumetric refractive index profiles, overcoming the design limitations of planar nanofabrication \cite{Wang2023,Schmid2019,Dottermusch2019}. In combination with structured light fields, refractive propulsion provides a controlled mechanism for optically driven particle motion based purely on momentum transfer, enabling systematic exploration of light-induced dynamics in microscale systems without reliance on absorption or thermally mediated effects.

\section{Methods}
\subsection{Equations of motion}
To formulate a dynamic model for the motion of an SBRIP particle, we first introduce the notion of generalized coordinates combining the translational and rotational degrees of freedom of the particle as
\begin{align}
\gvec{r}=\begin{pmatrix}\vec{r}\\
\vec{\phi}
\end{pmatrix}, \qquad \gvec{v}=\begin{pmatrix}\dot{\vec{r}}\\
\vec{\omega}
\end{pmatrix}, \qquad \gvec{F}=\begin{pmatrix}\vec{F}\\
\vec{\tau}
\end{pmatrix} ,
\end{align}
where $\vec{r} = (x,y,z)^\text{T}$ is the position of the particle, $\vec{\phi} = (\varphi, \theta, \chi)^\text{T}$ is the orientation of the particle expressed in Euler angles, $\vec{\omega}$ is the angular velocity of the particle, and $\vec{F}$ and $\vec{\tau}$ are the force and torque acting on the particle, respectively. Using these conventions, the equation of motion can be derived from the general overdamped Langevin equation for active Brownian particles \cite{WittkowskiL2012} as the special case without external forces and torques and without the drift term to
\begin{align}
\label{eq:eom}
\gvec{v}
=\beta\DD(\gvec{r})\RR^{-1}(\gvec{r})\big(\gvec{F}_{A}+\gvec{\xi}\big).
\end{align}
Here, $\RR(\gvec{r})$ is a $6\times 6$-dimensional transformation matrix that maps between the orientation of the laboratory frame and that of the particle. It is defined as the block-diagonal matrix
\begin{align}
    \RR(\gvec{r})=\begin{pmatrix}\mtx{R}(\vec{\phi}) & \Null\\
\Null & \mtx{R}(\vec{\phi})
\end{pmatrix}
\end{align}
with the $3 \times 3$-dimensional rotation matrix parameterized by the $z$-$y$-$z$ Euler angles
\begin{align}
\mtx{R}(\vec{\phi}) = \mtx{R}_{z}(\chi)\mtx{R}_{y}(\theta)\mtx{R}_{z}(\phi) ,
\end{align}
where the elementary rotation matrices about the $y$ and $z$ axes for an arbitrary angle $\theta$ are given by
\begin{align}
    \mtx{R}_{y}(\theta) &= \begin{pmatrix} \cos(\theta) & 0 & -\sin(\theta) \\ 0 & 1 & 0 \\ \sin(\theta) & 0 & \cos(\theta) \end{pmatrix}, \\
    \mtx{R}_{z}(\theta) &= \begin{pmatrix} \cos(\theta) & \sin(\theta) & 0 \\ -\sin(\theta) & \cos(\theta) & 0 \\ 0 & 0 & 1 \end{pmatrix}.
\end{align}
Furthermore, $\DD(\gvec{r})$ represents the generalized diffusion tensor, which is defined as
\begin{align}
    \DD(\gvec{r})=\frac{1}{\beta\eta}\,\RR^{-1}(\gvec{r})\,\HH^{-1}\,\RR(\gvec{r})
\end{align}
with the thermodynamic beta $\beta = (k_\text{B}T)^{-1}$, the Boltzmann constant $k_\text{B}$, the temperature $T$, the dynamic shear viscosity of the suspending medium (water) $\eta$, and the shape-dependent hydrodynamic resistance matrix $\HH$. The latter quantity relates the translational and angular velocity $\gvec{v}_\text{part}$ of a particle moving through an unbounded fluid medium at low Reynolds numbers with the drag force and torque $\gvec{F}_\text{drag}$ acting on the particle \cite{HappelB1991,WittkowskiL2012,voss2018hydrodynamic}
\begin{align}
\label{eq:drag_def}
    \gvec{F}_{\text{drag}}=-\eta\HH\,\gvec{v}_{\text{part}}.
\end{align}
The internal force $\vec{F}_A$ and torque $\vec{\tau}_A$ that propel the SBRIP particle are encoded as the active force-torque vector
\begin{align}
    \gvec{F}_{A}=\begin{pmatrix}\vec{F}_{A}\\
\vec{\tau}_{A}
\end{pmatrix}.
\end{align}
Finally, $\vec{\xi}(t)$ denotes the stochastic influence (Brownian motion) of the medium on the particle in the form of Gaussian white noise with the mean and variance defined by
\begin{align}
\big\langle \gvec{\xi}(t)\big\rangle &=\vec{0},\\
\big\langle \gvec{\xi}(t)\otimes\gvec{\xi}(t')\big\rangle &=\frac{2\eta}{\beta}\,\HH\,\delta(t-t') .
\end{align}

\subsection{Simulations}
\subsubsection{Simulation of the light propagation}
Since the particle dimensions are much larger than the illumination wavelength, we describe light propagation within the framework of geometrical optics and model the incident light field as an ensemble of rays. In a medium with a continuous refractive index profile $n(\vec{r})$, ray trajectories $\vec{r}(s)$ follow the Eikonal equation
\begin{equation}
\frac{\mathrm d}{\mathrm ds}\left(n(\vec{r})\,\frac{\mathrm d\vec{r}}{\mathrm ds}\right)=\Nabla n(\vec{r}),
\label{eq:ray_equation}
\end{equation}
where $s$ denotes the arc length along the ray path. In regions of constant refractive index ($\Nabla n=\vec{0}$), Eq.\ \eqref{eq:ray_equation} yields straight ray trajectories. At interfaces where $n(\vec{r})$ changes discontinuously, $\vec{r}(s)$ is not differentiable with respect to $s$ as the rays change direction due to refraction and specular reflection with the former being described by Snell's law. The transmission and reflectance probabilities are modeled by the Fresnel equations. In general, these probabilities depend on the polarization state of the incident ray. As there were no effects of polarization observed in the experiments, we neglect polarization by averaging the transmission and reflectance coefficients over all polarization states.
The optical force $\vec{F}_{A}$ on the particle is then obtained from momentum conservation by summing the momentum fluxes of all incident and outgoing (reflected and transmitted) ray components and taking their difference, which yields the net momentum transferred to the particle. Assuming a planar light source emitting with a homogeneous intensity $I$ in the normal direction of the plane, the total force transferred to an illuminated particle can be expressed as
\begin{equation}
\vec{F}_{A} = \frac{I}{c}\int_{S_{\text{L}}}\!\!\bigg(\hat{u}_{\text{in}}(\vec{r})-\!\!\!\sum_{\hat{u}_{\text{out}}\in O(\vec{r})}\kappa(\vec{r},\hat{u}_{\text{out}})\hat{u}_{\text{out}}\bigg)\mathrm{d}A ,
\label{eq:optical_force}
\end{equation}
where $c$ is the speed of light in the suspending medium, $S_\textrm{L}$ is the surface of the light source, $\hat{u}_\textrm{in}(\vec{r})$ is the unit vector in the direction of the light ray emitted at $\vec{r}$ by the light source, $O(\vec{r})$ is the set of every outgoing ray direction as unit vectors associated with the ray emitted at $\vec{r}$, $\kappa(\vec{r},\hat{u}_{\text{out}})$ is the probability that the ray emitted at $\vec{r}$ will exit with the direction $\hat{u}_{\text{out}}$, and $\mathrm{d}A$ is the differential surface element. Similarly, one can determine the total torque of the particle as
\begin{equation}
\begin{split}
\vec{\tau}_{A}=&\,\frac{I}{c}\int_{S_{\text{L}}}\!\!\bigg(\vec{r}\times\hat{u}_{\text{in}}(\vec{r}) \\
& -\!\!\!\sum_{\hat{u}_{\text{out}}\in O(\vec{r})}\kappa(\vec{r},\hat{u}_{\text{out}})(\vec{r}_{\text{out}}(\vec{r},\hat{u}_{\text{out}})\times\hat{u}_{\text{out}})\bigg)\mathrm{d}A
\end{split}
\label{eq:optical_torque}
\end{equation}
with $\vec{r}_{\text{out}}(\vec{r},\hat{u}_{\text{out}})$ being any point on the outgoing ray with direction $\hat{u}_{\text{out}}$ associated with a ray emitted by the light source at $\vec{r}$. 

Numerically, ray propagation and light-matter interaction were simulated using the GRINRAY framework \cite{Grinray}, a Monte-Carlo ray-tracing library designed for optical simulations including materials with a linear gradient index. Rays are propagated by alternating between (i) integrating Eq.~\eqref{eq:ray_equation} within each medium and (ii) applying the interface rules at refractive-index discontinuities. Particle geometries can be implemented by either specifying an exact solution for ray intersections or by representing the geometry as a signed distance function (SDF) $f(\vec{r})$. The absolute value of this function $\norm{f(\vec{r})}$ gives for every point $\vec{r}$ the distance to the surface of the object. If $\vec{r}$ is inside the geometry, $f(\vec{r})$ is negative, otherwise it is positive. The approach using SDFs utilizes a ray marching scheme to find intersections between a given ray and a particle geometry. This is less accurate than analytical ray intersection tests, but greatly simplifies the construction of more complex particle shapes. In the case of a material with a linear refractive-index gradient, an analytical solution to Eq.~\eqref{eq:ray_equation} is used to speed up the ray intersection test for both analytical and SDF-based geometries. Using GRINRAY, it is possible to sample the distribution of outgoing rays for a given incident ray. By discretizing the light source into a finite amount of ray bundles, the surface integrals in Eqs.\ \eqref{eq:optical_force} and \eqref{eq:optical_torque} can be approximately solved. In our simulations, a spatial discretization of $\Delta x = \Delta y = 0.1\;\textrm{\textmu m}$ and a ray-bundle size of 32 were used.

\subsubsection{Simulation of the particle motion}
To obtain the physical trajectories of the SBRIP particles, the stochastic equation of motion defined in Eq.\ \eqref{eq:eom} was integrated numerically. For this, we employed the Euler-Maruyama method to advance the system in discrete time steps $\Delta t$. At each iteration $i$, the generalized position vector $\gvec{r}$ is updated according to the discretized scheme
\begin{equation}
    \gvec{r}_{i+1} = \gvec{r}_i + \beta\DD(\gvec{r}_i)\RR^{-1}(\gvec{r}_i)\left(\gvec{F}_{A} \Delta t +  \left(\frac{2\eta\Delta t}{\beta} \mathcal{H}\right)^{\frac{1}{2}} \vec{X}_{N,i}\right)
    \label{eq:euler-maruyama-sbrip}
\end{equation}
with $\vec{X}_{N,i}$ being a vector of independent samples from a normal distribution with zero mean and variance $\sigma^2=1$. Here $\left(\mathcal{H}\right)^{1/2}$ denotes the matrix square root defined by $\left(\mathcal{H}\right)^{1/2} \left(\mathcal{H}\right)^{1/2} = \mathcal{H}$. The square root of $\mathcal{H}$ is uniquely defined as $\mathcal{H}$ is both symmetric and positive definite \cite{Brenner1967,Horn2013} and it can be calculated numerically by a Schur decomposition of $\mathcal{H}$ \cite{Higham1987}. In our simulations, a time step $\Delta t = 1/150\;\textrm{s}$ was used, which corresponds to one tenth of the framerate used to capture the experimental trajectories. We are not explicitly simulating the interaction between the substrate and the particle and instead rely on experimental observations of the steady-state tilt angles of particles undergoing motion and constrain the simulations to two spatial dimensions.

\subsubsection{Calculation of the hydrodynamic resistance matrix}
To determine the resistance matrix $\HH$ in Eq.\ \eqref{eq:drag_def}, we compute the hydrodynamic force and torque resulting from prescribed translational and rotational motions of the particle in Stokes flow. Specifically, six simulations are performed: three unit translations along the Cartesian basis vectors and three unit rotations about these axes. In each case, the force and torque are obtained by integrating the surface traction over the particle boundary.

The Stokes equations are solved numerically using the computational fluid dynamics software AcoDyn \cite{acodyn}. 
They are given by 
\begin{align}
    \eta \Laplace\vec{u} - \Nabla p &= \vec{0} , \\
    \Nabla\cdot\vec{u} &= 0 ,
\end{align}
where $\vec{u}$ and $p$ denote the fluid velocity and pressure fields, respectively.
The fluid simulations were performed on unstructured tetrahedral finite-volume grids with locally varying mesh resolution and typically $\approx 2.5\times 10^6$ cells, depending on the particle shape.
A cubic domain of edge length \SI{400}{\micro\metre} centered around the particle was found sufficient to suppress boundary effects.

We write the hydrodynamic resistance matrix in $3\times 3$ block form as
\begin{equation}
\HH=
\begin{pmatrix}
\KK & \CC^{\mathrm{T}} \\
\CC & \OO
\end{pmatrix},
\label{eq:H_block}
\end{equation}
where $\KK$ and $\OO$ are the translational and rotational resistance tensors, and $\CC$ is the translation-rotation coupling tensor. Torques are taken about the particle center of mass. With this choice, $\HH$ is symmetric within numerical accuracy. Components that vanish by symmetry were set to zero and the absolute values of components that have to be identical by symmetry were averaged.  
For validation and comparison, the hydrodynamic resistance matrix was also calculated with the software package Hydrosub \cite{Garcia2002}. While AcoDyn provides higher fidelity by explicitly modeling the particle geometry and not relying on approximate analytic equations, the translation-rotation coupling tensor $\CC$ for the hemisphere exhibited high numerical noise in the solutions. Consequently, for this specific case, we utilized the tensor $\CC$ obtained from Hydrosub. While Hydrosub employs significant approximations regarding the particle shape, this value proved to be more significant for our calculations.
\begin{widetext}
\noindent\begin{minipage}{\textwidth}
\noindent Our results for the blocks $\KK$, $\CC$, and $\OO$ of the hydrodynamic resistance matrix for the cap, cone, cornet, hemisphere, and sphere particles are (for symmetry axis along the $z$ axis) given by 
\begin{align}
\KK_\mathrm{cap}&=\diag(77.19,77.19,87.68)\,\unit{\micro\meter},& 
\CC_\mathrm{cap}&=\begin{pmatrix} 0      & -45.01  &   0\\
45.01  &   0     &   0\\
0      &   0     &   0
\end{pmatrix}\unit{\micro\square\meter} ,&
\OO_\mathrm{cap}&=\diag(1967,1967,2279)\,\unit{\micro\meter\cubed} ,
\label{eq:cap}\raisetag{3ex}
\\
\KK_\mathrm{cone}&=\diag(66.03,66.03,80.23)\,\unit{\micro\meter},& 
\CC_\mathrm{cone}&=\begin{pmatrix} 0      & -50.65  &   0\\
50.65  &   0     &   0\\
0      &   0     &   0
\end{pmatrix}\unit{\micro\square\meter} ,&
\OO_\mathrm{cone}&=\diag(1498,1498,1592)\,\unit{\micro\meter\cubed} ,
\label{eq:cone}\raisetag{3ex}
\\
\KK_\mathrm{cor}&=\diag(65.70,65.70,80.22)\,\unit{\micro\meter},& 
\CC_\mathrm{cor}&=\begin{pmatrix} 0      & -70.65  &   0\\
70.65  &   0     &   0\\
0      &   0     &   0
\end{pmatrix}\unit{\micro\square\meter} ,&
\OO_\mathrm{cor}&=\diag(1534,1534,1588)\,\unit{\micro\meter\cubed} ,
\label{eq:cornet}\raisetag{3ex}
\\
\KK_\mathrm{hemi}&=\diag(77.48,77.48,87.64)\,\unit{\micro\meter},& 
\CC_\mathrm{hemi}&=\begin{pmatrix} 0      & -18.55 & 0 \\
18.55  &   0     &   0\\
0      &   0     &   0
\end{pmatrix}\unit{\micro\square\meter} ,&
\OO_\mathrm{hemi}&=\diag(1944,1944,2280)\,\unit{\micro\meter\cubed} ,
\label{eq:hemisphere}\raisetag{3ex}
\\
\KK_\mathrm{sphere}&=75.71\,\unit{\micro\meter}\,\Eins ,& 
\CC_\mathrm{sphere}&=\Null ,&
\OO_\mathrm{sphere}&=1583\,\unit{\micro\meter\cubed}\,\Eins 
\label{eq:sphere}
\end{align}
\end{minipage}
\end{widetext}
with the zero matrix $\Null$ and the identity matrix $\Eins$.

\subsection{Experimental methods}
\subsubsection{Two-photon polymerization}
Two-photon polymerization (TPP) of the SBRIP particles is carried out in a custom-built setup using a pulsed femtosecond laser (FemtoFiber Pro, Toptica) operating at a central wavelength of \SI{780}{\nano\metre} and a repetition rate of \SI{80}{\mega\hertz}. The beam is expanded and directed via a dichroic mirror into a 100$\times$ oil immersion objective (CFI Plan Apo $\lambda$D, $\text{NA} = 1.45$, Nikon), which tightly focuses the radiation into the photoresist.

Three-dimensional structuring is achieved by translating the sample with a piezoelectric stage (P-563.3CD, Physik Instrumente) with a travel range of \SI{300}{\micro\metre} in all spatial directions and a resolution of \SI{2}{\nano\metre}. The writing speed is adjustable up to \SI[per-mode = symbol]{100}{\micro\metre\per\second} for typical scan distances. For larger lateral displacements, additional stepper stages (M-229.25S, Physik Instrumente) extend the accessible writing area in the $x$-$y$ plane.

The mean laser power at the sample is dynamically controlled using an acousto-optic modulator and typically in the range \qtyrange[range-phrase = --]{0.3}{1}{\milli\watt}. Under these conditions, lateral voxel sizes below \SI{500}{\nano\metre} and longitudinal voxel sizes below \SI{1}{\micro\metre} are achieved. The fabrication process is monitored in situ via CMOS imaging under LED illumination filtered with a yellow long-pass filter to suppress unintended exposure.

Precise alignment of the writing plane with the substrate surface is ensured by coupling a continuous-wave \SI{532}{\nano\metre} laser into the beam path. Here, the back reflection off the substrate is used for autofocus routine. This alignment step is critical to guarantee adhesion of the polymerized structures to the substrate and to prevent deformation due to detachment during writing.

All devices used in the fabrication process are controlled with custom Matlab scripts.

\subsubsection{3D printing of SBRIP particles}
\begin{figure*}
    \centering
    \includegraphics[width=\linewidth]{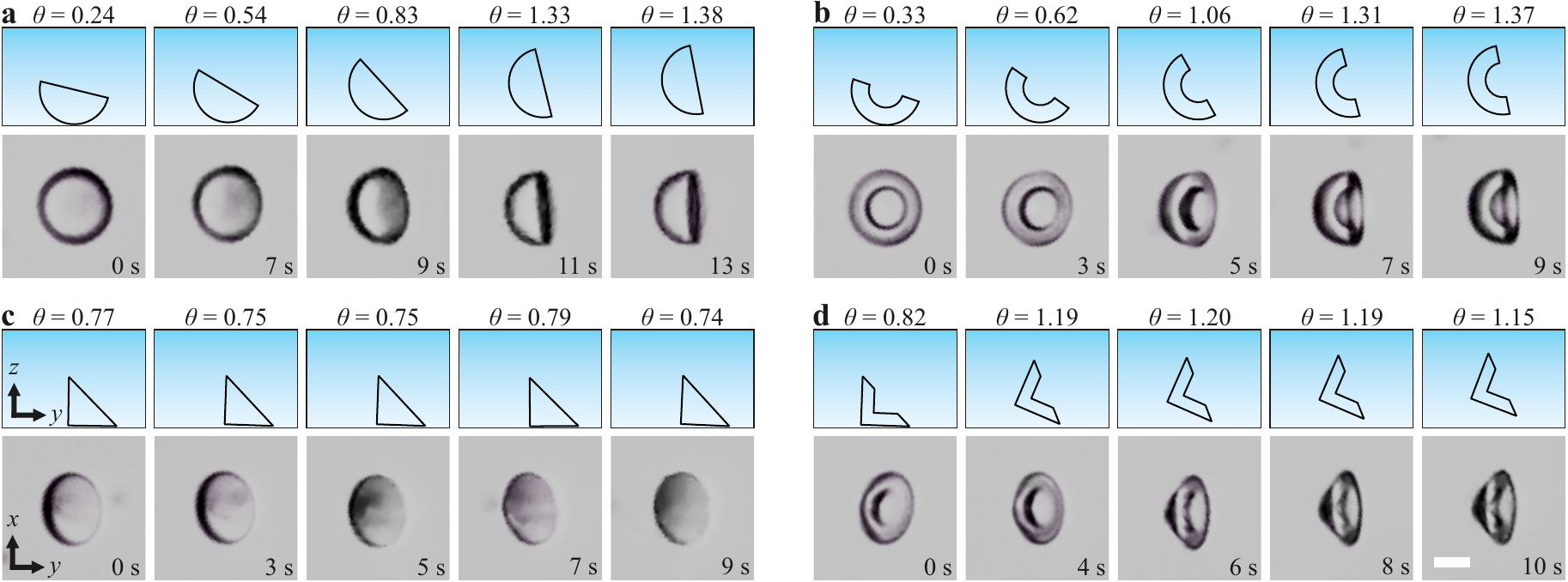}
    \caption{Reorientation of SBRIP particles under illumination from below. Top row: schematic side views illustrating the reorientation of \textbf{a} hemispheres, \textbf{b} caps, \textbf{c} cones, and \textbf{d} cornets at different time points with measured tilt angle on top. In addition, the estimated upward movement of the hemispheres, caps, and cornets is shown schematically. Bottom row: corresponding microscopy images recorded from below. Scale bar: \SI{5}{\micro\meter}.}
    \label{fig:02}
\end{figure*}
SBRIP particles with the geometries shown in Fig.~\ref{fig:01}\textbf{a} are 3D printed, including spheres (radius $r=\SI{4}{\micro\metre}$), hemispheres ($r=\SI{5}{\micro\metre}$), cones ($r=\SI{5}{\micro\metre}$, height $h=\SI{5}{\micro\metre}$), caps ($r=\SI{5}{\micro\metre}$, radius of the spherical cutout $r_{\text{cut}}=\SI{3}{\micro\metre}$), and cornets ($r=\SI{5}{\micro\metre}$, $h=\SI{5}{\micro\metre}$, radius $r_{\text{cut}}=\SI{3}{\micro\metre}$ and height $h_{\text{cut}}=\SI{3}{\micro\metre}$ of the conical cutout). The reduced radius of the spherical particles is chosen to approximately match the volume of the hemispherical geometry. The 3D models of these particles are designed in a CAD program (FreeCAD) and the corresponding writing path is calculated via a slicer software (Slic3r).

All structures are printed from the negative-tone photoresist OrmoComp (micro resist technology GmbH), selected for its mechanical stability and optical transparency in the visible range. A droplet of the resin (\SI{\approx 100}{\micro\litre}) is deposited onto a standard glass coverslip (size 1, $\SI{22}{\milli\metre}\times\SI{22}{\milli\metre}$, Marienfeld) and heated to \SI{80}{\degreeCelsius} for \SI{5}{\minute} to remove trapped air, and subsequently mounted in the TPP setup.

For particles with a homogeneous refractive index profile, the mean writing laser power at the sample plane is set to \SI{0.65}{\milli\watt}. For gradient particles, the laser power is varied uniformly along the $z$-axis during fabrication, spanning a range from \qtyrange{0.45}{0.77}{\milli\watt}. In all cases, the writing velocity is fixed at \SI[per-mode = symbol]{100}{\micro\meter\per\second}. Depending on particle size and geometry, fabrication of a single structure requires approximately \qtyrange[range-phrase = --]{3}{8}{\minute}.

Particles are written sequentially in a rectangular array with center-to-center distances of \SI{50}{\micro\metre} in both lateral directions, using the integrated stepper stages for large-area positioning. During fabrication, adhesion to the substrate is required to prevent displacement of the polymerizing structure. To enable controlled release after processing, predefined breaking points in the form of conical supports are incorporated beneath each particle. The broad base of the cone ensures stable attachment during writing, while the narrow tip connects to the particle body.

Following fabrication, unpolymerized photoresist is removed by immersing the sample three times in developer solution (OrmoDev, micro resist technology) for \SI{7}{\minute} each, followed by thorough rinsing with isopropanol. After complete drying, a fluid chamber is assembled around the particle array using an adhesive spacer (\SI{120}{\micro\metre} height, \SI{9}{\milli\metre} diameter opening, Grace Bio-Labs SecureSeal). The chamber is filled with deionized water and sealed with a second coverslip. Final detachment of the particles from the support structures is achieved by exposure to an ultrasonic bath (S40H, frequency of \SI{40}{\kilo\hertz}, Palssonic) for around \SI{5}{\second}.

\subsubsection{Particle propulsion}
Particle propulsion experiments are conducted using an inverted microscope (Eclipse Ti, Nikon) coupled to a continuous-wave near-infrared laser (Smart Laser Systems) operating at \SI{1064}{\nano\metre} with an output power of up to \SI{2.5}{\watt}. The Gaussian laser beam is enlarged, collimated, coupled into the microscope, and directed through a 20$\times$ objective (CFI Plan Fluor 20$\times$, $\text{NA} = 0.5$, Nikon), resulting in nearly homogeneous illumination of the sample from below.
At the sample plane, the illuminated region forms a circular area with a diameter of approximately \SI{300}{\micro\metre}, corresponding to an average intensity of \SI{1.17e7}{\watt\per\square\metre}. 
Particle dynamics are recorded in real time using a CCD camera with a frame rate of 15 frames per second and subsequently analyzed by particle tracking with the TrackMate plugin (Simple LAP Tracker, ImageJ). From the tracked trajectories, particle positions and velocities are extracted. Symmetry-broken particles can exhibit different orientations during propulsion, which can be defined as the angle between the incident light direction and the symmetry axis of the particle. This angle can be determined by measuring the seen eccentricity of the circular flat sides of the particles and then fitting it with a skew normal distribution using maximum likelihood estimation. The probability density function (PDF) of a skew normal distribution is defined as
\begin{equation}
    P_{\textrm{skew}}(\theta)=P_{\textrm{normal}}\Big( \frac{\theta-\Lambda}{\varsigma} \Big) \bigg( 1 + \textrm{erf}\bigg(\alpha \frac{\theta-\Lambda}{\sqrt{2}\varsigma}\bigg)\! \bigg) ,
    \label{eq:skewnorm}
\end{equation}
where $P_{\textrm{normal}}$ is the PDF of a normal distribution, $\textrm{erf}$ is the error function, and $\Lambda$, $\varsigma$, and $\alpha$ are the location, scale, and shape parameters of the skew normal distribution, respectively.

\section{Results}
We distinguish between two classes of SBRIP particles based on the origin of the symmetry breaking: shape-symmetry-broken particles, where the optical symmetry breaking arises from the refractive contrast at the particle boundary, and gradient-index-profile (GRIN) particles, where it is generated by internal refractive-index gradients.

\subsection{Shape-symmetry-broken particles}
\subsubsection{Particle propulsion}
\begin{table*}
    \caption{Parameters for fitting a skew normal distribution to the experimentally observed angle $\theta$ of moving SBRIP particles with different shapes. Also listed are the averages of the tilt angle $\theta$, the lateral force $F_y$, the normal force $F_z$ and the tilting torque $\tau_x$ calculated from 10,000 samples of the aforementioned distribution. Also shown are the average particle speeds obtained via simulations $\langle \norm{\vec{v}} \rangle_{\textrm{sim}}$ and observed experimentally $\langle \norm{\vec{v}} \rangle_{\textrm{exp}}$ for each shape.}
    \label{tab:01}
    \begin{tabular}{|c|c|c|c|c|c|c|c|c|c|}
    \hline 
     & \multicolumn{3}{c|}{Fit parameters} &  \multicolumn{4}{c|}{Averages for fitted distribution} & \multicolumn{2}{c|}{Average velocity} \tabularnewline
    \hline 
    Shape & $\Lambda/\textrm{rad}$ & $\varsigma/\textrm{rad}$ & $\alpha$ & $\left\langle \theta\right\rangle /\textrm{rad}$& $\left\langle F_{y}\right\rangle /\text{pN}$ & $\left\langle F_{z}\right\rangle /\text{pN}$ & $\left\langle \tau_{x}\right\rangle / (\text{pN}\,\text{\textmu m})$&$\langle \norm{\vec{v}} \rangle_{\textrm{exp}} / (\textrm{\textmu m} \, \textrm{s}^{-1})$&$\langle \norm{\vec{v}} \rangle_{\textrm{sim}} / (\textrm{\textmu m} \, \textrm{s}^{-1})$\tabularnewline
    \hline 
    \hline 
    Hemisphere & $1.295$ & $0.05106$ & $1.285$ & $1.327$ & $-0.3163$ & $0.3301$ & $0.2034$ & $1.69\pm0.48$ & $3.78\pm0.05$ \tabularnewline
    \hline 
    Cap & $1.345$ & $0.1710$ & $-8.556$ & $1.209$ & $-0.3531$ & $0.3844$ & $-0.6057$ & $3.85\pm0.46$ & $4.12\pm0.05$ \tabularnewline
    \hline 
    Cone & $0.8661$ & $0.06516$ & $-1.301$ & $0.8249$ & $-0.2901$ & $0.08711$ & $0.2207$ & $2.28\pm0.28$ & $4.13\pm0.06$ \tabularnewline
    \hline 
    Cornet & $1.272$ & $0.09347$  & $-2.780$ & $1.202$ & $-0.3079$ & $0.3529$ & $-1.362$ & $3.69\pm0.41$ & $3.81\pm0.06$ \tabularnewline
    \hline 
    \end{tabular}
\end{table*}

\begin{figure*}
    \centering
    \includegraphics[width=\linewidth]{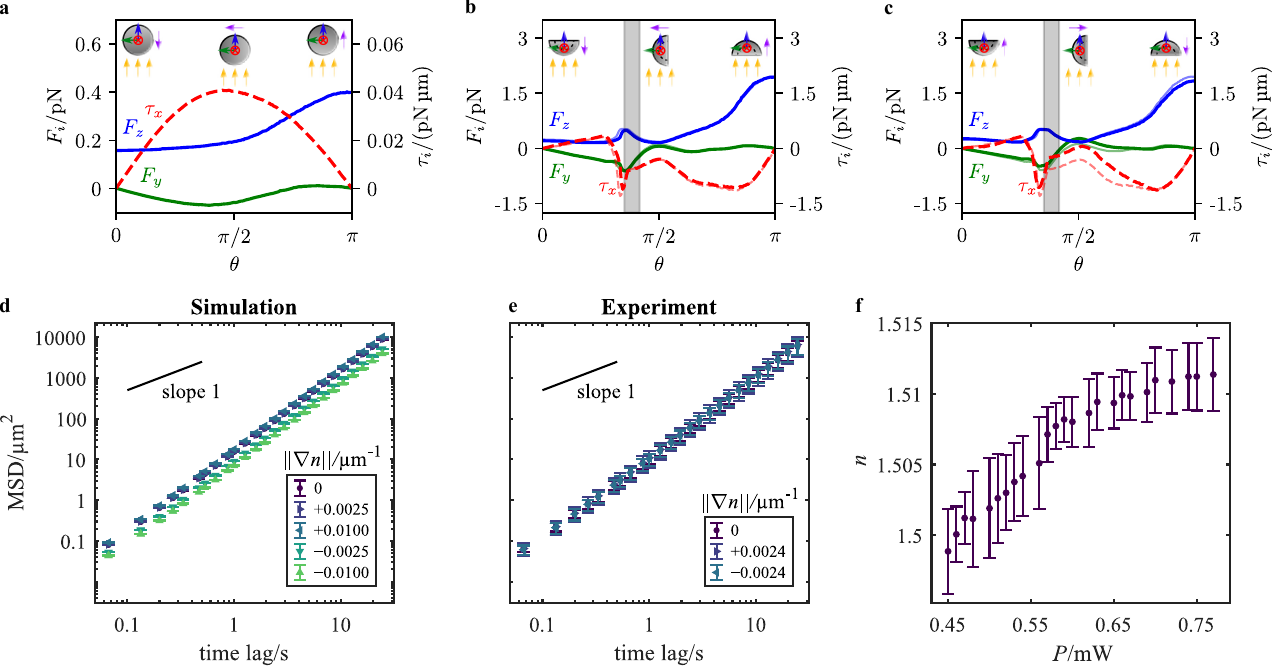}
    \caption{Forces and torque acting on \textbf{a} spheres and \textbf{b}, \textbf{c} cap-shaped particles with a linear GRIN profile \textbf{b} parallel to the symmetry axis and rising towards the round side of the particle and \textbf{c} going in the opposite direction. The GRIN strength in subfigures \textbf{a}, \textbf{b}, and \textbf{c} is $\norm{\nabla n} = \SI{0.01}{\per\micro\metre}$ and the particles are illuminated from $z=-\infty$ in the direction of the $z$-axis. At the center of the spherical particle and at half-height of the cap-shaped particles, the refractive index is fixed at $n=1.505$. The angle between the incident light and the symmetry axis of the particle is denoted as $\theta$. The purple arrow shows the direction of the refractive-index gradient with the arrow head pointing towards higher refractive indices. Highlighted in grey is the central 68.3\% confidence interval of the angles observed experimentally. This is omitted in subfigure \textbf{a} as the orientation of a sphere cannot be determined by visual inspection. Shown in subfigures \textbf{b} and \textbf{c} with thinner lines and a paler color scheme are the forces and torque acting on an equivalent cap-shaped particle without a refractive-index gradient. \textbf{d}, \textbf{e} MSD curves of cap-shaped particles with refractive-index gradients with varying strength and directionality in \textbf{d} simulation and \textbf{e} experiment. \textbf{f} Resulting refractive index (at $\lambda=\SI{532}{\nano\metre}$) of OrmoComp depending on the mean laser power during two-photon polymerization.}
    \label{fig:06}
\end{figure*}

To evaluate the propulsion capabilities of SBRIP particles, we investigated the dynamics of spheres, hemispheres, caps, cones, and cornets as five distinct geometries (Fig.~\ref{fig:01}\textbf{a}) both experimentally and numerically. For the experiments, particles of these shapes were fabricated via two-photon polymerization (Fig.~\ref{fig:01}\textbf{c}). Here, the writing power is set to \SI{0.65}{\milli\watt}. This results in a refractive index of $n=\SI{1.509}{}$ at \SI{532}{\nano\meter} (see Fig.~\ref{fig:06}\textbf{f}), which was measured with digital holographic tomography (HT-2H, Tomocube). After fabrication and development, a fluid chamber was assembled around the particle array, filled with deionized water, and sealed with a second coverslip. The particles were then released from their sacrificial support structures by brief exposure to an ultrasonic bath.

After detachment, the particles sedimented onto the glass substrate, where they adopted stable resting configurations determined by their geometry. Hemispheres and bowl-shaped particles (caps and cornets) rested either with the curved side or with the opening facing the substrate, corresponding to two distinct resting orientations. Cones, in contrast, predominantly settled on their lateral surface, remaining in contact with the substrate along that side. 
The motion of the different particles was subsequently observed under homogeneous illumination from below (Fig.~\ref{fig:01}\textbf{d}).

\begin{figure*}
    \centering
    \includegraphics[width=\linewidth]{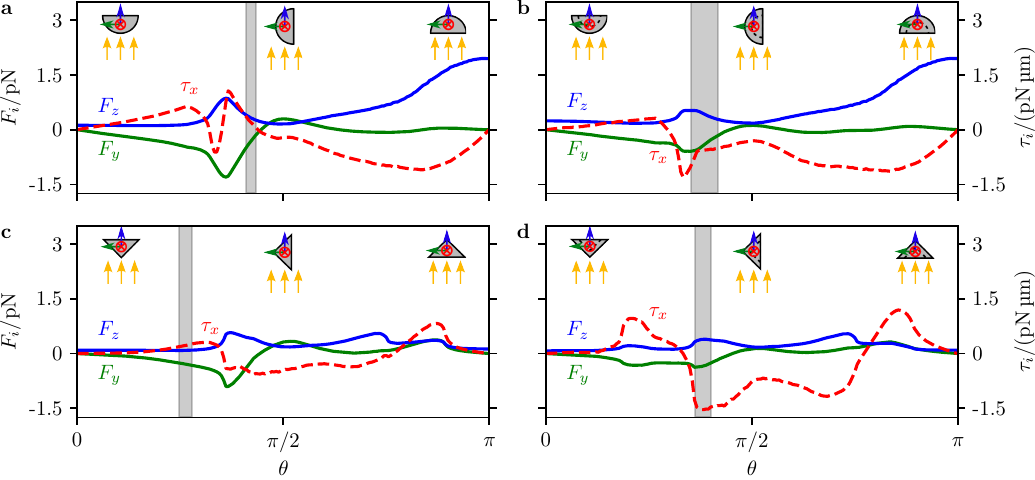}
    \caption{Simulation results for the forces and torques acting on particles with the shape of a \textbf{a} hemisphere, \textbf{b} spherical cap, \textbf{c} cone, and \textbf{d} cornet. The incident light rays start at $z=-\infty$ and propagate in parallel to the positive $z$-axis. The angle $\theta$ denotes the angle between the incident light direction and the symmetry axis of the particle. Forces are displayed with a solid line, while torques are shown with a dashed line. For reasons of symmetry, there are no torques aligned to the $y$- and $z$-axis and no force in the direction of the $x$-axis. Shown in grey is the central 68.3\% confidence interval of the angles observed experimentally for moving particles.} 
    \label{fig:03}
\end{figure*}
Upon the illumination, the SBRIP particles experienced an upward optical force and a shape-dependent optical torque, which caused them to reorient relative to the substrate. The reorientation dynamics depended strongly on the particle geometry: hemispheres, caps, and cornets rotated from their initial resting configurations toward more upright orientations, whereas cones largely remained on their side and rotated only slightly during illumination. At the same time, the particles began to translate laterally along the substrate under the action of the optical forces. We first focus on the reorientation dynamics. Figure~\ref{fig:02} presents representative rotation sequences together with schematic side views illustrating the corresponding particle orientations during illumination (see also Supplementary Video~1).

Due to the rotational symmetry of the SBRIP particles, their orientation can be described by a single tilt angle $\theta$ between the symmetry axis of the particle and the incident light. For a particle with the curved side facing downward and the flat side parallel to the substrate, the angle is defined as $\theta=0$, whereas the inverted configuration with the curved side facing upward corresponds to $\theta=\pi$.

Because Brownian motion causes fluctuations of the tilt angle during propulsion, the orientation is not constant in time. We therefore approximate the experimentally observed tilt-angle distribution by a skew normal distribution of $n=150$ measurements and sample tilt angles from this distribution (Tab.~\ref{tab:01}). Cones remain at an angle close to $\pi/4$, corresponding approximately to their opening angle. 

To understand the origin of the observed particle motion, we compute the optical forces and torques acting on the different SBRIP particle geometries using ray-tracing simulations. The incident light field is modeled as a planar source emitting parallel rays in the positive $z$ direction with homogeneous intensity. The particle orientation is varied by changing the tilt angle $\theta$ between the illumination direction and the particle symmetry axis.

\begin{figure*}
    \centering
    \includegraphics[width=\linewidth]{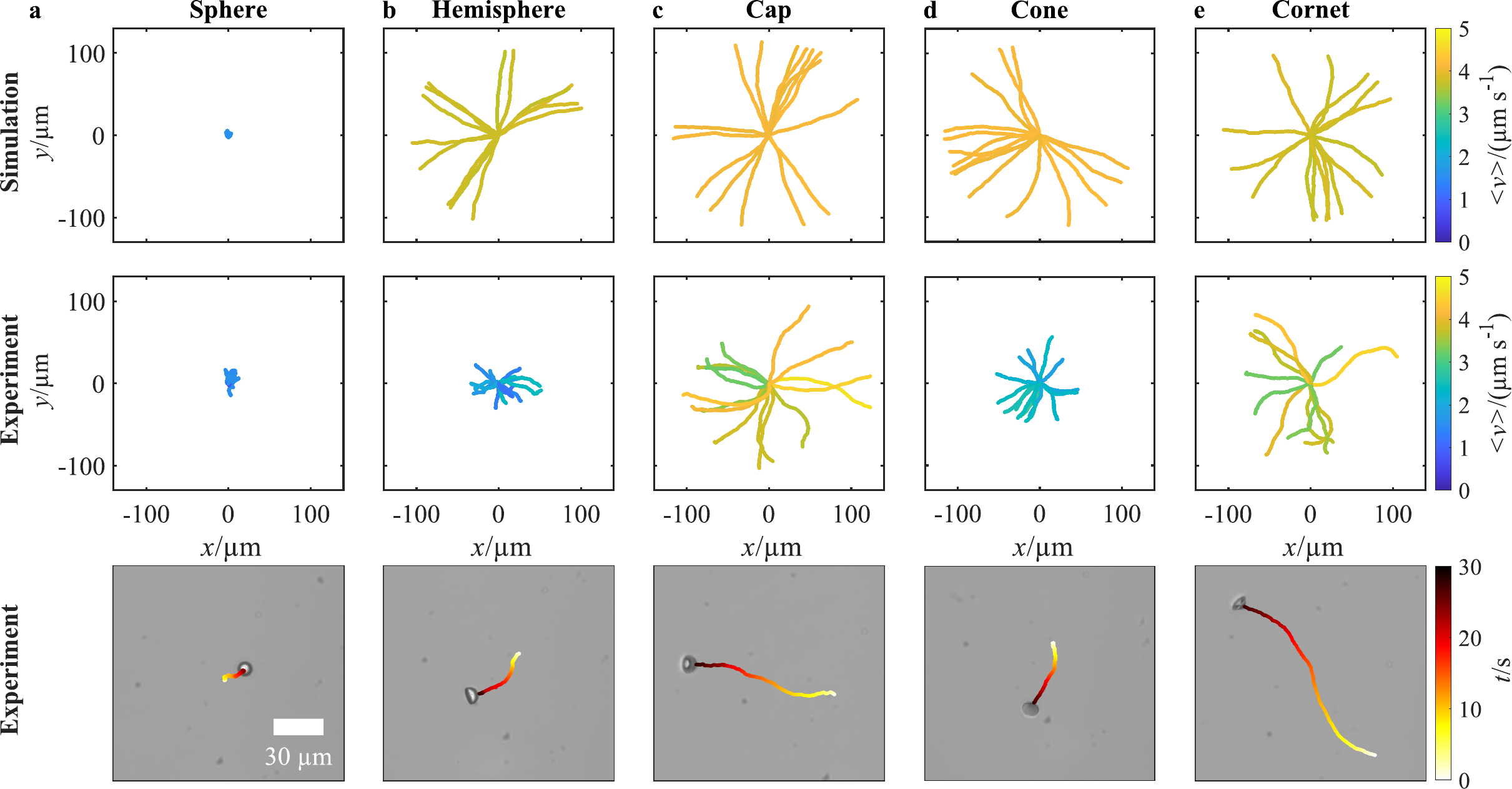}
    \caption{Lateral trajectories of propelled \textbf{a} spheres, \textbf{b} hemispheres, \textbf{c} caps, \textbf{d} cones, and \textbf{e} cornets in the first \SI{30}{\second} of illumination after their initial reorientation. The first row shows the simulated and the second row the corresponding experimental trajectories with the average speed along each trajectory shown by coloration. In the third row, one experimentally measured trajectory for each shape is overlaid on top of a microscope image showing the state of the particle at $t=\SI{30}{\second}$.}
    \label{fig:04}
\end{figure*}

Ray trajectories are calculated using the Eikonal equation within each medium and Snell's law together with Fresnel coefficients at refractive-index discontinuities. The optical force and torque are then obtained from momentum conservation by summing the momentum fluxes of all incident, reflected, and transmitted ray components and integrating over the discretized light source. Repeating this procedure for different tilt angles yields the angular dependence of the force and torque.

Because the investigated particle geometries possess rotational symmetry around their main axis, the optical response is constrained by symmetry. In particular, if the axis of symmetry of the particle lies in the $y$-$z$ plane, the force component along the $x$-direction vanishes and no torques occur about the $y$- or $z$-axes. For the orientations $\theta=0$ and $\theta=\pi$, where the illumination direction coincides with the particle symmetry axis, both the lateral force $F_y$ and the tilting torque $\tau_x$ vanish as well for reasons of symmetry.

The calculated lateral force $F_y(\theta)$ and tilting torque $\tau_x(\theta)$ for hemispheres, caps, cones, and cornets as functions of the tilt angle $\theta$ are presented in Fig.~\ref{fig:03}. Both quantities depend strongly on the particle orientation. While the tilting torque $\tau_x$ determines the particle orientation under illumination, the lateral force component $F_y$ drives propulsion along the substrate. Consistent with the experimental observations, the simulations reveal a pronounced dependence of the torque on the tilt angle, which explains the reorientation dynamics observed experimentally (Fig.~\ref{fig:02}). At the same time, the lateral force also varies strongly with $\theta$, implying that the propulsion strength depends sensitively on the instantaneous particle orientation.

A comparison of the different particle geometries reveals systematic differences in the angular dependence of both $F_y(\theta)$ and $\tau_x(\theta)$. In particular, the cutout geometries (caps and cornets) exhibit a reduced lateral force compared to their closed counterparts. At the same time, the change in the angular dependence of the tilting torque is more pronounced as the torque at the tilt angle observed experimentally has opposite signs when comparing hemisphere and cap or cone and cornet, respectively.

This behaviour can be traced back to reflections occurring at the planar surface of the particles. Rays incident on this boundary reverse their momentum component along the surface normal, which contributes significantly to the force and torque acting on the particle. For geometries with a cutout, a large fraction of this planar interface is removed, reducing this contribution and thereby altering the resulting optical response.

\subsubsection{Trajectories}
\begin{figure*}
    \centering
    \includegraphics[width=\linewidth]{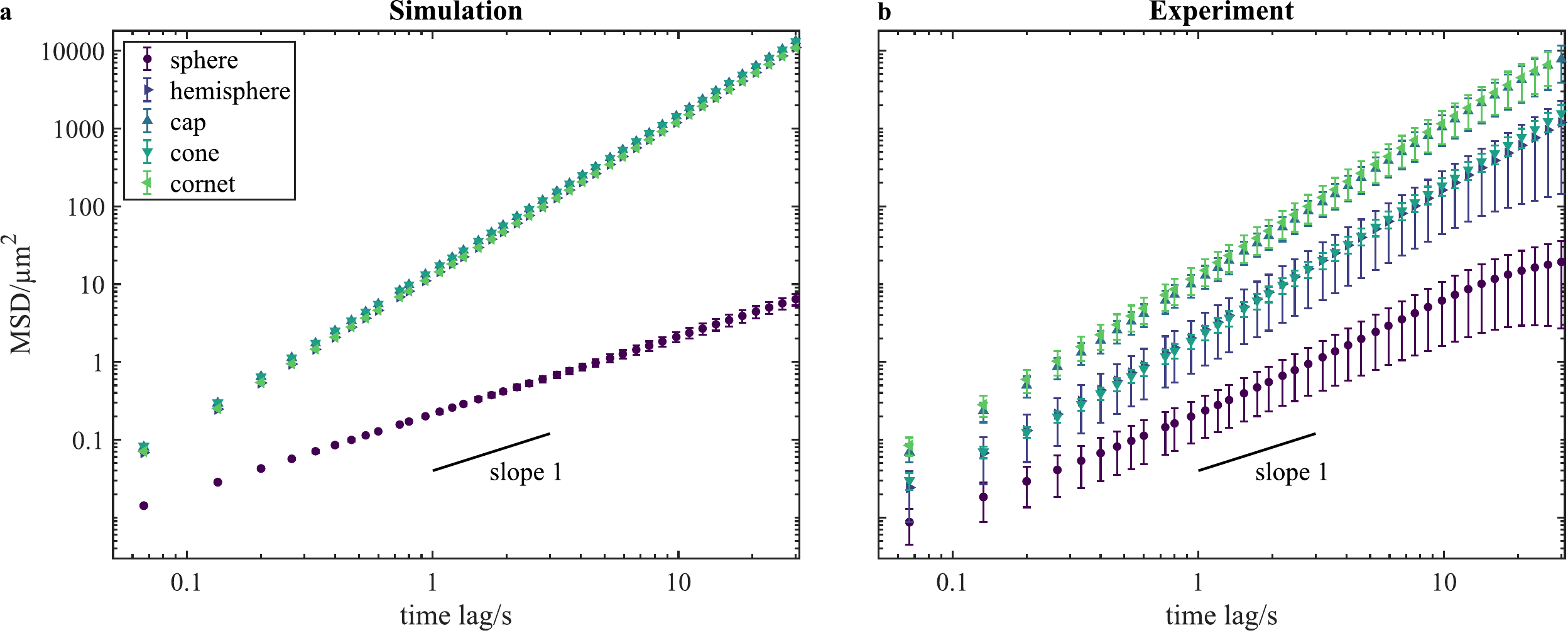}
    \caption{Mean squared displacement of propelled microparticles (sphere, hemisphere, cap, cone, and cornet) as a function of time lag in \textbf{a} simulation and \textbf{b} experiment. The error bars depict the standard deviation. Additionally, the black bar denotes the unity slope which corresponds to purely diffusive motion.}
    \label{fig:05}
\end{figure*}
To characterize the particle motion under illumination, we analyze both experimental and simulated trajectories of SBRIP particles. In the experiments, individual particle trajectories are obtained from video microscopy, while in the simulations the particle motion is computed using the optical forces derived in the previous section together with the hydrodynamic resistance matrices given in Eqs.~\eqref{eq:cap}-\eqref{eq:sphere}. The resulting equations of motion are integrated using the scheme described in Eq.~\eqref{eq:euler-maruyama-sbrip}. Representative trajectories from experiments and simulations are compared in Fig.~\ref{fig:04}, while the corresponding mean squared displacements (MSD) are shown in Fig.~\ref{fig:05}. In the experimental results, trajectories and MSD curves are evaluated only after the initial reorientation phase following the onset of illumination. This is reflected in simulations by assuming a constant tilt angle $\theta$.

In the idealized ray-tracing model, perfectly spherical particles do not experience a deterministic lateral propulsion force under homogeneous illumination, and their simulated motion is therefore  diffusive (Fig.~\ref{fig:04}\textbf{a}, Fig.~\ref{fig:05}\textbf{a}). Experimentally, nominally spherical particles can nevertheless show weak directed motion, which we attribute to fabrication-induced deviations from perfect sphericity that break the symmetry and generate a finite lateral force (Fig.~\ref{fig:04}\textbf{a}, Supplementary Video~2). Consistent with this, the MSD remains close to diffusive scaling but can exhibit a small superdiffusive component for individual particles (Fig.~\ref{fig:05}\textbf{b}).

In contrast, introducing geometric symmetry breaking leads to finite lateral optical forces and torques, resulting in persistent particle motion in both simulations and experiments. As shown in Fig.~\ref{fig:05}, symmetry-broken particle shapes exhibit MSD curves with slopes approaching two, corresponding to the ballistic regime of active particle motion. In addition to lateral propulsion, hemispheres, caps, and cornets also display a small vertical displacement component, consistent with the comparatively large normal forces predicted for these particle shapes (Tab.~\ref{tab:01}).

For the cutout geometries (caps and cornets), simulations and experiments show good qualitative agreement (Fig.~\ref{fig:04}\textbf{c} and \textbf{e}, Supplementary Videos~5 and 6). For hemispheres and cones, however, the simulated trajectories are noticeably longer than those observed experimentally (Fig.~\ref{fig:04}\textbf{b} and \textbf{d}, Supplementary Videos~3 and 4), reflecting higher average propulsion speeds in the simulations (Tab.~\ref{tab:01}).

For hemispheres, we attribute this deviation primarily to the simplified treatment of particle orientation in the simulations. Experimentally, hemispherical particles frequently exhibit a stop-and-go behavior in which the particle alternates between phases of active propulsion and temporary rest at slightly different tilt angles. Because the simulations assume a constant tilt angle during propulsion, these intermittent resting phases are not captured, which leads to an overestimation of the effective propulsion velocity.

For cone particles, the deviation likely originates from hydrodynamic interactions with the nearby substrate. In the simulations, the hydrodynamic resistance matrices are calculated for particles moving in an unbounded fluid. In the experiments, however, the particles move in close proximity to the substrate, which increases the hydrodynamic resistance. This effect is expected to be particularly pronounced for cones, which maintain contact with the substrate along their lateral surface due to their small tilt angle. As a result, the simulations likely underestimate the hydrodynamic drag experienced by cones and therefore overestimate their propulsion speed.
Additional discrepancies between simulations and experiments may also arise from fabrication imperfections that lead to slight deviations from the ideal particle geometries assumed in the simulations as was seen in the case of spherical particles.

\subsection{Gradient-index-profile particles}

\subsubsection{Polymer-based particles}

To demonstrate the possibility of particle propulsion using only a symmetry-breaking refractive-index gradient, we consider a spherical particle with an axial GRIN profile but no geometric symmetry breaking. The angular dependence of the corresponding optical forces and torque obtained from ray-tracing simulations is shown in Fig.\ \ref{fig:06}\textbf{a}. Much like for shape-symmetry-broken particles, the particle experiences a finite normal force $F_z$ in the direction of the incident light for all orientations. In addition, a transverse force component $F_y$ emerges for configurations where the refractive-index gradient is not (anti-)parallel to the incident light. This shows that a refractive-index gradient alone can generate lateral optical forces even in the absence of shape asymmetry.

However, due to the angular dependence of the optical torque $\tau_x$, it is unlikely that this particle shape is viable for particle propulsion. In the simulations, $\tau_x$ vanishes only at the symmetry-aligned orientations $\varphi=0$ and $\varphi=\pi$, where $F_y$ is also (approximately) zero. For intermediate tilt angles where $F_y$ becomes appreciable, the torque remains consistently positive over $0<\varphi<\pi$, causing a monotonic reorientation toward a torque-free state as a spherical particle cannot generate a compensating torque from the contact with the substrate. Consequently, orientations that exhibit finite transverse force are not mechanically stable, and any lateral drift is expected to be transient as the particle rotates toward an orientation with $F_y \approx 0$. 

To test the effect of a GRIN profile on particle propulsion we therefore examined cap-shaped particles with a GRIN profile. In particular, we tested particles with a GRIN profile that is parallel to the particle orientation, i.e., a rising refractive index towards the rounded end of the particle, and particles with a GRIN profile that is antiparallel to the orientation, i.e., a rising index towards the end of the particle where the cutout is located.

Simulation results for a gradient strength of $\norm{\nabla n} = \SI{0.01}{\per\micro\metre}$ (Fig.\ \ref{fig:06}\textbf{b} and \textbf{c}) show that a refractive-index gradient parallel to the particle orientation slightly enhances the forces and torques acting on the particle. However, this effect remains very small within the angular range corresponding to stable motion. In the antiparallel configuration, propulsion is suppressed compared to the case without a GRIN profile. However, this result should be considered indicative at best as the introduction of this gradient profile also appears to enhance the inaccuracies inherent in SDF raytracing in GRINRAY thus leading to a relatively high number of rays that could not be fully traced.

GRIN particles can be fabricated by modulating the writing intensity during TPP (see Methods). In this case, laser powers in a range from \qtyrange{0.45}{0.77}{\milli\watt} lead to refractive index values of \qtyrange{1.499}{1.511}{} at $\lambda=\SI{532}{\nano\metre}$ (see Fig.~\ref{fig:06}\textbf{f}), as determined using digital holographic tomography (HT-2H, Tomocube). Lower laser powers result in an incomplete polymerization, while higher powers cause thermal damage and bubble formation in the fabrication process, which limits the fabrication window. Consequently, cap-shaped particles with refractive-index gradients parallel and antiparallel to the particle orientation were fabricated with gradient strengths of $\norm{\nabla n} = \SI{\pm0.0024}{\per\micro\metre}$ (from \qtyrange{1.499}{1.511}{} over \SI{5}{\micro\metre}).

The MSDs of cap particles with an (anti-)parallel GRIN profile (see Fig.\ \ref{fig:06}\textbf{e}, Supplementary Videos~7 and 8) show that this gradient strength does not lead to a noticeable increase in propulsion as the MSD of particles with a gradient index fall within the error range of that of a homogeneous particle. The simulation results in Fig.\ \ref{fig:06}\textbf{d} match this observation for the case of the parallel gradient direction and show that even a five-fold increase in gradient strength, which is far beyond the capabilities of the material used for particle fabrication here, would not lead to a significant change in propulsion speed. For the antiparallel case, the MSD grows slightly slower, however, the fact that this shift is seemingly independent of the gradient strength and cannot be confirmed experimentally corroborates the hypothesis above that this is a numerical artifact due to the loss of rays in the simulation.

\subsubsection{Silicon-based particles}
While we were not able to construct a particle without a symmetry-breaking in the shape of the particle from Ormocomp that is continuously moving in the experiments, numerical simulations suggest that such a feat should be possible using Silicon-based particles with a photonic metastructure that acts as a linear refractive index gradient. By utilizing subwavelength gratings, it is possible to tune the refractive index between that of Silicon ($n\approx3.5$) and that of Silicon oxide ($n\approx1.45$) \cite{LuqueGonzalez2021}. Since spherical particles are infeasible for continuous propulsion, we chose a cubic particle with an edge length of \SI{10}{\micro\metre} and a mean refractive index of $2.5$. The gradient index was applied perpendicular to two of the faces of the particle with a strength of \SI{0.1}{\per\micro\metre}. Under illumination orthogonal to the gradient direction, our simulations yield a lateral force of \SI{0.3094}{\pico\newton}, which is comparable to the lateral forces for particles with a symmetry-broken shape (c.f.\ Table \ref{tab:01}).

\section{Conclusions}
We have introduced a class of 3D-printable light-driven microparticles based on a symmetry-broken refractive index distribution, where direct momentum transfer from symmetry-broken light refraction enables sustained self-propulsion. The required symmetry breaking can be achieved both by employing symmetry-breaking particle shapes, such as hemispheres or cones, or gradient index materials. Using SBRIP particles, it is thus possible to realize active matter without the drawbacks of absorption-based propulsion schemes or chemically fueled systems where increased thermal dissipation, low penetration in bulk media, and fuel depletion cause deviations from the ideal model of Active Brownian particles.

In our work, we have analyzed SBRIP particles from both a theoretical perspective and with an experimental approach. On one hand, we have developed a model for this class of particles that can be solved numerically with a combination of raytracing and hydrodynamic simulations. On the other hand, we demonstrated in experiments the use of two-photon polymerization as a method of fabrication for these particles with few design constraints. Both approaches complement each other as experiments offer access to quantities such as the tilt angle of the particles which are difficult to predict theoretically, while simulations allow us to peer outside of the experimentally accessible range of parameters, such as in the case of the refractive-index gradient strength. We have also identified the hydrodynamic interaction between the particles and the substrate as a key influence on the particle motion by comparing experimental and simulation results.

For the future, we expect SBRIP particles to become a robust and versatile platform for light-actuated active matter that grants the high degree of controllability in optical systems while also allowing for activity in bulk suspensions. In particular, we look forward to applications of dense suspensions of SBRIP particles where the combination of strong hydrodynamic interactions between particles and optical feedback allows for the design of adaptive material systems.

\section{Data Availability}
The raw data and plot scripts for the figures as well as the recorded videos of particles under illumination are available at Zenodo \cite{ancillary_files_jeggle_rueschenbaum2026}.

\section{Code Availability}
The raytracing code GRINRAY is available at Zenodo \cite{Grinray}. The simulation code for particle trajectories is also available at Zenodo \cite{ancillary_files_jeggle_rueschenbaum2026}.

\section{Acknowledgements}
We thank Marc Beuel, Niklas Vollmar, and Daniel Wendland for helpful discussions.  
This work is funded by the Deutsche Forschungsgemeinschaft (DFG, German Research Foundation) -- 433682494 -- SFB 1459; SFB 1459/2 2025 -- 433682494.
R.W.\ is funded by the Deutsche Forschungsgemeinschaft (DFG, German Research Foundation) -- 535275785.
We gratefully acknowledge \'{A}lvaro Barroso, Bj\"orn Kemper, and Anne Marzi from the Biomedical Technology Center of the Medical Faculty, University of M\"unster, for supporting the refractive index measurements of the microparticles with the 3D microscopic imaging device Tomocube HT-2H.

\section{Author Contributions}
A.P.\ calculated the hydrodynamic resistance matrices with AcoDyn. 
I.K.\ calculated the hydrodynamic resistance matrices with Hydrosub. 
J.J.\ performed the raytracing simulations, calculated the trajectories, and performed the other theoretical work.
M.R\"u.\ designed and evaluated the experiments. E.V.\ performed the fabrication and tracking of particles. J.S.\ made the scanning electron microscope images. 
J.I. co-supervised M.R\"u.\ and E.V..
C.D.\ supervised M.R\"u.\ and E.V..
M.Re.\ supervised J.S.. 
R.W.\ supervised J.J., A.P., and I.K..
C.D., M.Re., and R.W.\ supervised the project.
C.D.\ and R.W.\ conceived the project. 
R.W.\ conceptualized the SBRIP particles.
All authors contributed to discussion, scientific presentation, and manuscript writing.

\section{Additional information}
\textbf{Competing interests}: The authors declare no competing interests.

\bibliographystyle{naturemag}
\bibliography{refs}

\end{document}